\documentclass[acmlarge]{acmart}
\makeatletter
\newcommand{\myconfshort}{\acmConference@shortname}
\newcommand{\myconffull}{\acmConference@name}
\newcommand{\myconfdate}{\acmConference@date}
\newcommand{\myconfloc}{\acmConference@venue}
\AtBeginDocument{
  \fancypagestyle{firstpagestyle}{
    \fancyhead{}%
    \fancyfoot[C]{}%
  }
  \fancyhf{}
  \fancyhead[LO]{\@headfootfont\shorttitle}%
  \fancyhead[RE]{\@headfootfont\@shortauthors}%
  \fancyhead[LE]{\@headfootfont\footnotesize \myconfshort, \myconfdate, \myconfloc}%
  \fancyhead[RO]{\@headfootfont\footnotesize \myconfshort, \myconfdate, \myconfloc}%
  \fancyfoot[C]{}%
}
\makeatother
\acmBooktitle{\conffull\@ (\confshort), \confdate, \confloc}
\AtBeginDocument{%
  }

\copyrightyear{2026}
\acmYear{2026}
\setcopyright{cc}
\setcctype{by}
\acmConference[FAccT '26]{The 2026 ACM Conference on Fairness, Accountability, and Transparency}{June 25--28, 2026}{Montreal, QC, Canada}
\acmBooktitle{The 2026 ACM Conference on Fairness, Accountability, and Transparency (FAccT '26), June 25--28, 2026, Montreal, QC, Canada}
\acmDOI{10.1145/3805689.3806723}
\acmISBN{979-8-4007-2596-8/2026/06}

\usepackage{array}
\usepackage{color-edits}
\usepackage{tabularx}
\usepackage{ragged2e}
\usepackage{pdflscape}
\addauthor[Ningjing]{suppress}{violet}

\begin{document}

\title[Beyond the Single Turn]{Beyond the Single Turn: Reframing Refusals as Dynamic Experiences Embedded in the Context of Mental Health Support Interactions with LLMs}


\author{Ningjing Tang}
\affiliation{%
  \institution{Carnegie Mellon University}
  \city{Pittsburgh}
  \state{Pennsylvania}
  \country{USA}
}
\author{Alice Qian}
\affiliation{%
  \institution{Carnegie Mellon University}
  \city{Pittsburgh}
  \state{Pennsylvania}
  \country{USA}
}
\authornote{Co-second authors contributed equally to this work.}
\author{Qiaosi Wang}
\affiliation{%
  \institution{Carnegie Mellon University}
  \city{Pittsburgh}
  \state{Pennsylvania}
  \country{USA}
}
\authornotemark[1]
\author{Esther Howe}
\affiliation{
\institution{University of Washington School of Medicine}
\country{USA}
}
\author{Blake Bullwinkel}
\affiliation{%
  \institution{Microsoft}
  \city{Redmond}
  \country{USA}
}
\author{Paola Pedrelli}
\affiliation{
\institution{Harvard Medical School}
\city{Boston}
\country{USA}
}
\author{Jina Suh}
\affiliation{
\institution{University of Washington}
\city{Seattle}
\country{USA}
}
\author{Hoda Heidari}
\affiliation{%
  \institution{Carnegie Mellon University}
  \city{Pittsburgh}
  \state{Pennsylvania}
  \country{USA}
}
\authornote{Co-last authors contributed equally to this work.}
\author{Hong Shen}
\affiliation{%
  \institution{Carnegie Mellon University}
  \city{Pittsburgh}
  \state{Pennsylvania}
  \country{USA}
}
\authornotemark[2]
\renewcommand{\shortauthors}{Tang et al.}

\begin{abstract}
\textit{Content Warning: This paper contains participant quotes and discussions related to mental health challenges, emotional distress, and suicidal ideation.}

Large language models (LLMs) are increasingly used for mental health support, yet the model safeguards---particularly refusals to engage with sensitive content---remain poorly understood from the perspectives of users and mental health professionals (MHPs) and have been reported to cause real-world harms. This paper presents findings from a sequential mixed-methods study examining how LLM refusals are experienced and interpreted in mental health support interactions. Through surveys (N=53) and in-depth interviews (N=16) with individuals using LLMs for mental health support and MHPs, we reveal that refusals are not isolated, single-turn system behaviors but rather constitute dynamic, multi-phase experiences: pre-refusal expectation formation, refusal triggering and encounter, refusal message framing, resource referral provision, and post-refusal outcomes. We contribute a multi-phase framework for evaluating refusals beyond binary policy compliance accuracy and design recommendations for future refusal mechanisms. These findings suggest that understanding LLM refusals requires moving beyond single-turn interactions toward recognizing them as holistic experiences embedded within users' support-seeking trajectories and the broader LLM design pipeline.
\end{abstract}





\maketitle


\section{Introduction}

Large Language Models (LLMs) are increasingly utilized by end-users for mental health support, valued for their accessibility and availability across a spectrum of needs ranging from emotional processing to acute distress~\cite{Song2024-iq, Siddals2024-xx}. However, this usage has raised significant ethical and safety concerns, including risks of misinformation and psychological harm, evidenced by real world incidents including user deaths, and prompting recent legislative action~\cite{Moore2025-rh, Hill2025-ln, CaliforniaSB243, IllinoisHB1806}. Concurrently, AI developers have implemented a variety of model safeguards that restrict or redirect model behavior when user requests are deemed risky~\cite{OpenAI2025-uv, OpenAI2023-ts, Yuan2025-sd, Guan2024-jd}.

One of the most common strategies for safeguarding models from generating potentially harmful outputs 
is via a \emph{refusal}: when an LLM system explicitly or implicitly declines to engage with a user’s request in a way that meets the user's contextual expectation \cite{Xie2024-jh}. Prior literature on AI safety has primarily framed refusals as a technical optimization problem, focusing on detecting high-risk inputs and triggering appropriate non-compliance responses~\cite{Xie2024-jh}. These approaches typically evaluate refusals through benchmarks constitutes single-turn conversations that measure whether the system correctly refuses certain requests based on accuracy metrics~\cite{Xie2024-jh, Vidgen2023-vd}.  Despite the promise of refusals as a safety instrument, recent empirical work has begun to document refusal-like interactions that are experienced as a form of ``denial of service," producing feelings of abandonment, invalidation, or harm for vulnerable users seeking mental health support 
~\cite{chandra2025lived, Siddals2024-xx}. Unlike many other emergent LLM failure modes (e.g., hallucination), refusals are \emph{deliberately designed safety instruments} intended to protect users in sensitive contexts. Yet we still lack a comprehensive, stakeholder-grounded understanding of how refusals actually unfold and are experienced by users seeking mental health support. This understanding is necessary to make refusal guardrails accountable to their intended purpose: reducing harm in practice, not only satisfying benchmarked notions of AI safety.

Understanding the impact of LLM safety guardrails, such as refusals, on users seeking mental health support requires attending to multiple forms of expertise. Individuals with lived experience in seeking mental health support bring ``experiential expertise'' \cite{young2019toward} from their everyday interactions with LLM systems. Mental Health Professionals (MHPs), in contrast, bring domain expertise grounded in clinical practice. These perspectives are shaped by different responsibilities: users are accountable to lived consequences, while clinicians are accountable to “do no harm” and duty-of-care norms. Yet refusal mechanisms are typically designed within AI labs and product teams, with limited input from those who directly experience or clinically interpret their effects. Bringing these forms of expertise together is therefore crucial for examining refusal mechanisms and reasoning about their harms, benefits, and trade-offs.

Our study investigates how LLM refusals are experienced, interpreted, and reimagined in mental health contexts through these complementary, multi stakeholder perspectives. In particular,

\begin{itemize}
\item \textbf{RQ1}: How do \emph{end-users} experience refusals from LLMs when seeking mental health support?
\item \textbf{RQ2}: How do \textit{mental health professionals} perceive LLM refusals in instances when users are seeking mental health support?
\item \textbf{RQ3}: What \textit{design recommendations} do users and mental health professionals propose to address potential negative effects of refusals while maintaining appropriate boundaries between users and LLMs?
\end{itemize}

\label{fig:findingsframing}
\begin{figure*}[t]
  \centering
  \includegraphics[width=\textwidth]{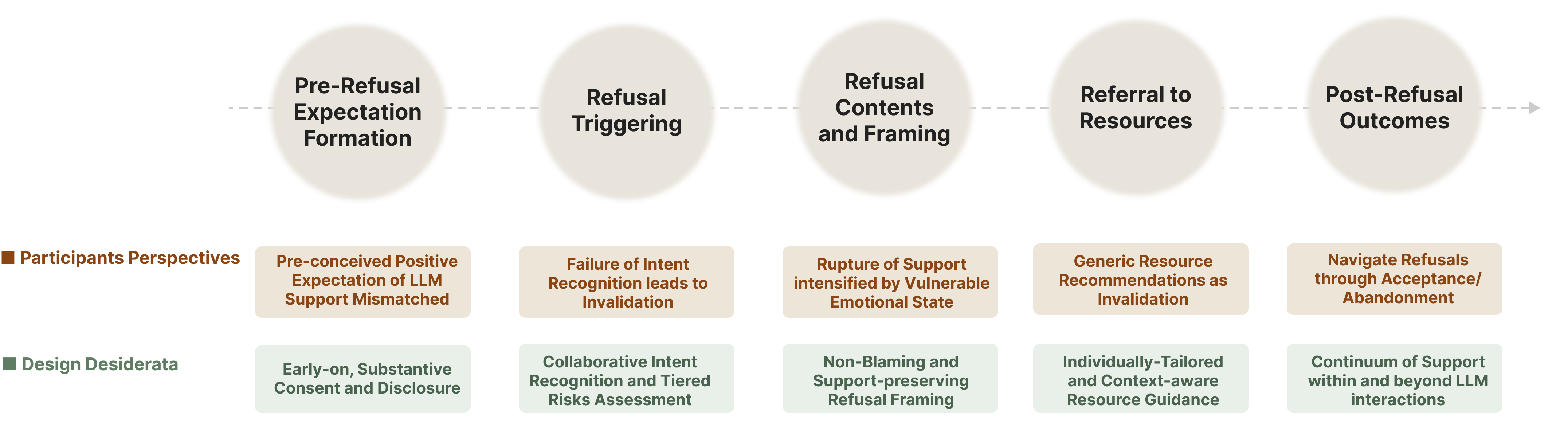}
  \Description{This image shows 5 different circles that represent different stages of user experience. 5 circles are aligned horizontally from left to right.}
  \caption{\textbf{Our proposed multi-phase framework of understanding LLM refusals in mental health support interactions as dynamic experiences. This framing is the result of 53 surveys and 16 interviews with end-users and mental health professionals. The framework reveals how refusal experiences unfold in phases including expectation formation, intent recognition, refusal framing, resource provision, and post-refusal outcomes. } 
 }
  \label{fig:multi-phase-refusal}
\end{figure*}



To answer these questions, we conducted a sequential mixed-methods study \cite{Ivankova2006-id} involving individuals who had lived experience using LLMs for mental health support and Mental Health Professionals (MHP). We first conducted an exploratory survey (N=53), followed by 
semi-structured interviews (N=16). Our findings reveal that refusals are best understood not as isolated single-turn outputs but as \emph{dynamic, multi-phase experiences} that start prior to LLM interactions and evolve over the course of the interactions with an LLM. Drawing on user accounts and MHP interpretations, we show how harms and benefits emerge across five stages: (1) pre-refusal expectation formation, (2) refusal triggering and encounter, (3) refusal message framing, (4) resource referral provision, and (5) post-refusal outcomes. 

\looseness=-1

Building on these insights, we argue that examining LLM safeguard in mental health interactions requires moving beyond single-turn refusal accuracy toward a more holistic understanding of refusals as an experiential trajectory. 
We identified a set of design recommendations from our participants that map on to the multiple stages of refusals, including proactive and substantive disclosure before refusal happens; collaborative and tiered intent recognition; support-preserving refusal framing; tailored, context-aware resource guidance; and post-refusal continuum of support within and beyond LLM interactions.

In summary, we contribute (1) an empirical account of how users and mental health professionals perceive and interpret LLM refusals in mental health interactions, (2) a multi-phase framework for analyzing refusal-related opportunities and harms (illustrated in Fig~\ref{fig:findingsframing}), and (3) design considerations for refusal mechanisms that integrate clinical expertise while accounting for lived experience. Together, these contributions offer guidance for designing refusal behaviors in LLMs that maintain appropriate boundaries without compounding distress or undermining pathways to care. 

\section{Background and Related Works}
\subsection{Refusals as LLM Behavior for Safety and Value Alignment}
Refusals are a core mechanism for enforcing safety and value alignment in language models \cite{OpenAI2023-ts, Han2024-xr, Guan2024-cu}. They are commonly defined as intentional noncompliance with user requests due to safety constraints, policy violations, or contextual inappropriateness \cite{Xie2024-jh, Brahman2024-hk, Foundational_Contributors2024-xh}. Prior work distinguishes epistemic abstention \cite{Wen2025-jt}, as well as “technical” and “social” refusals \cite{Wester2024-fm}. We define LLM refusals as cases in which a model fails, either explicitly or implicitly, to engage with a request in a way that satisfies contextual expectations, including both ``hard refusals'' and “safety completion” style responses that redirect the conversation to a different topic \cite{Guan2024-cu}.

Most existing research treats refusal as a desirable outcome of alignment optimization, focusing on mechanisms for detecting harmful inputs and triggering noncompliance \cite{Chao2024-wq, Xie2024-jh, Mazeika2024-pz, Zou2023-ks, Wang2024-yc, Vidgen2023-vd}. Various benchmarks have been developed to measure the ``safety'' of LLMs based on these refusal behaviors \cite{Chao2024-wq, Mazeika2024-pz, souly2024strongrejectjailbreaks, Xie2024-jh, Zou2023-ks}. However, they generally lack explicit policy rationales and therefore encode only vague or implicit notions of harm. Further, existing benchmarks are usually limited to single-turn examples, restricting their ability to capture more realistic and context-dependent scenarios \cite{Xie2024-jh, souly2024strongrejectjailbreaks}. Some recent approaches have incorporated more detailed safety policies into training and evaluation, including Constitutional AI \cite{Bai2022-gf} and Deliberative Alignment \cite{Guan2024-jd}. However, these methods continue to rely on LLMs, themselves, to make complex decisions about whether and how to comply with a given request. The widespread adoption of judge LLMs in scaling these decisions has raised concerns about their validity and reliability \cite{Chehbouni2025-nd, Weidinger2025-wm, Wallach2025-se}.

While the technical approaches to LLM refusal have been instrumental in scaling safety interventions, they often reflect a “thin” conception of alignment that prioritizes formal policy compliance over situated meaning \cite{Geertz1973-iv, Nelson2023-iy}. Recent scholarship has therefore argued for broader, context-sensitive accounts of alignment that attend to social and institutional dynamics \cite{Lazar2023-so, Weidinger2023-od, Selbst2019-fn, Joyce2021-ow, Shelby2023-lj}. Despite this shift, the ways in which refusals are experienced and interpreted in practice remains underexamined.

\subsection{LLMs for Mental Health Support: User Experience and Risks}

In the HCI research community, there is a long history of examining how end-users engage with digital technologies for mental health support \cite{De-Choudhury2014-qh, Pendse2021-yv}. This line of work has explored user experiences with social media platforms \cite{De-Choudhury2014-qh, Milton2023-xx}, rule-based chatbots \cite{Koulouri2022-ib}, and other types digital mental health technologies \cite{Haque2022-yb}, consistently demonstrating that these technologies create both opportunities for access and connection as well as new forms of risk and harm, particularly for people from marginalized communities \cite{Tanni2024-ee, Schaadhardt2023-iu}.

 Recent large-scale analyses show users increasingly turn to LLM-based chatbots for both clinical mental health issues and everyday emotional support \cite{Jung2025-nn}, drawn by perceived non-judgmental tones, accessibility, and anonymity \cite{Song2024-iq, Jung2025-nn, Siddals2024-xx, Li2025-fb, Zheng2025-jk, Yoo2025-fh}. Simultaneously, RAI researchers have raised significant ethical concerns \cite{Moore2025-rh, Iftikhar2025-wu, Chandra2024-wz, Tavory2024-vf, Rahsepar-Meadi2025-lm}. FAccT research demonstrates that LLM responses often misalign with therapists' ethical principles, producing misleading, stigmatizing, or emotionally harmful outputs \cite{Moore2025-rh, Iftikhar2025-wu}, while complementary work documents psychological harms from users' lived experiences \cite{Chandra2024-wz, Siddals2024-xx}. These challenges contributed to growing discussions around \emph{socioaffective alignment}, emphasizing the need for AI systems not only to be factually correct, but also to respond in ways that are relationally and emotionally attuned \cite{Kirk2025-zl} and to evaluate for and mitigate interactional harms~\cite{ibrahim2025towards}. 

Recent work documents refusal-like interactions as "denial of service" that can produce feelings of abandonment, invalidation, or harm 
\cite{Chandra2024-wz, Siddals2024-xx}. However, refusals appear only as one problematic behavior among many, not theorized as a distinct sociotechnical mechanism. This gap is significant because refusals are \emph{deliberately designed safety instruments}, unlike emergent or unintended LLM behaviors. While refusals can theoretically protect users, 
mental health interactions with technology are deeply relational, often occurring during distress, dependence, or crisis \cite{Ajmani2025-ry,Kirk2025-zl}. Refusals thus become a critical site for examining how intended safety mechanisms are manifested in lived impact. \looseness = -1

\section{Methods}
In this research, we focus on how LLM refusal behaviors are experienced and interpreted in mental health by users and MHPs through a sequential explanatory mixed-methods design \cite{Ivankova2006-id}. Given that LLM refusal is currently underexplored from a human-centered perspective, we first collected survey data to obtain a broad, preliminary understanding of user and MHP's perspectives on LLM refusal. Informed by these survey findings, we then conducted semi-structured interviews with both users and MHPs to gain deeper insights into their experiences, perspectives, and future visions. 


\subsection{Phase 1: Survey Study (N = 53)}
The survey was intended to capture broad user experiences and MHP perspectives on LLM refusals, collect real-world refusal cases, and screen for potential interview participants. We described our recruitment procedure in detail in Appendix~\ref{app:survey-methods}. All participants (N=53) were located in the United States, with 32 end-user participants and 21 MHP participants. User participants ranged in age from 18-66+, with 12 identifying as women, 17 as men, and 3 as non-binary/other. Participants reported using LLMs for various mental health support needs, including psychotherapy (n=23), interpersonal advice (n=15), coaching (n=22), and companionship (n=17). MHP participants included licensed therapists and counselors (n=9), psychiatrists (n=4), licensed psychologists (n=4), peer support specialists (n=2), and community mental health workers (n=2). 14 identifying as women, 6 as men, and 1 as non-binary/other.



\noindent \textit{\textbf{Survey Design.}}
The online Qualtrics survey had seven sections. 
After consenting and filling out background information, participants proceeded to role-specific questions. End-users completed questions about their LLM usage patterns, including frequency of use, specific platforms utilized, and types of mental health support sought based on the taxonomy provided by \cite{McCain2025-gs}. End-users were then asked about their refusal experiences, with those who had encountered refusals providing detailed open-text narratives describing the context, their intentions, the refusal they received, and their emotional reactions. MHPs were asked about their professional background, familiarity with LLMs, and whether they had experienced or heard of scenarios where clients encountered LLM refusals when seeking mental health support. \footnote{We emphasized that they should only share information within the bounds of client confidentiality. }

We then presented participants with refusal scenarios derived from refusal instances in mental health documented in news media and Reddit discussions (see Appendix~\ref{app:surveyscenariodevelopment} for detailed development methodology 
\footnote{While our systematic search used diverse keywords to capture refusal instances reported online, the resulting corpus predominantly contained instances that contains negative experiences. This reflects well-documented biases in online health discussions, where users disproportionately share problematic experiences seeking support or validation \cite{De-Choudhury2014-qh}. Satisfactory interactions often pass undocumented \cite{Golder2017-mn}. We acknowledge this limitation for our initial scenario dataset.}), 
following a similar data collection approach in prior work 
\cite{Gao2025-fu, Kawakami2024-mp}. End-users received scenarios aligned with their reported mental health support use, while MHPs were randomly assigned scenarios to capture diverse clinical perspectives. Both groups assessed the psychological impact of each refusal scenario they were assigned, rating the short-term and long-term psychological impact of the refusal on a 7-point scale (ranging from ``Extremely Harmful'' to ``Extremely Helpful'') and providing open-ended explanations for their rating.
\footnote{We define short-term and long-term impact as follow in the survey: 'Short-term' refers to the immediate period after the interaction, such as within a few minutes or the first few hours within the same day. 'Long-term' refers to a sustained impact on mental well-being over an extended period, such as days, weeks, months, or longer after the interaction.} The survey concluded with questions about participants' ideal refusal strategies, demographics, and interest in follow-up interviews. The survey remained open for two weeks in August 2025, 
with each participant receiving \$5 as compensation. 
Appendix~\ref{app:survey-methods} contains detailed survey structure.

\noindent \textit{\textbf{Survey Data Analysis.}}
For quantitative data, we calculated descriptive statistics via python.
For qualitative data analysis, we conducted reflexive thematic analysis \cite{Braun2006-vi}, 
with the first author conducting initial coding and the research team meeting weekly to develop higher-order themes.

\subsection{Phase 2: Interview Study (N = 16)}

We then conducted semi-structured interviews with 6 end-users (4 men, 2 women; ages 26-66+) and 10 mental health professionals (9 women, 1 man; ages 18-55+) recruited from survey respondents who signed up for follow-up studies (See full demographic details in Table~\ref{tab:interview-demographics} in Appendix~\ref{app:interview-methods}.). We used purposeful sampling \cite{Palinkas2015-yv} to intentionally recruit users seeking different type of help and professionals with diverse backgrounds. End-user participants reported using LLMs for companionship, psychotherapy, interpersonal advice, and life coaching. MHP participants included licensed therapists, psychologists, peer support specialists, and community mental health workers. 
Interviews  were conducted via zoom, lasting 40-60 minutes, with each participant receiving \$60 through Prolific or Amazon gift cards.

\noindent \textbf{Interview with Users.}
We began by inviting them to share their experience with using LLMs for mental health support. We asked them to walk through their refusal experiences at their own comfort level, exploring the types of conversations they typically had with LLM systems, the specific moments of refusal, their emotional responses, and what they wished had happened instead. 

\noindent \textbf{Interview with MHPs.}
We utilized 6 refusal scenarios collected from Phase 1 as probes for interviews with MHPs \cite{Jenkins2010-fr} (see Appendix~\ref{app:scenarioselection} for the scenario selection criteria). For MHPs, the interview structure focused on MHPs' interpretations of user-reported or MHP-reported refusal scenarios. 
After asking their professional background, we presented an overview of the 6 scenarios and asked professionals to pick 2-3 scenarios where they find the most interesting to discuss. For each scenario, we provided the user/MHP's own description of their experience, then engaged professionals in detailed discussions about whether they believed the AI responded appropriately, what specifically went wrong or right in the interaction, and what they considered as appropriate response from their standpoints. This approach allowed us to capture professional assessments while maintaining connection to genuine user experiences.

For both groups, we provided a set of six design recommendations derived from survey analysis to ground participants' answers in concrete examples~\cite{harrington2019deconstructing} and scaffold discussion of alternative refusal strategies. 


\noindent \textit{\textbf{Interview Data Analysis.}}
We employed reflexive thematic analysis to analyze interview data \cite{braun2012thematic,braun2019reflecting}. Open coding was carried out by the first author across the full corpus with analytical memoing, 
while the second and third authors provided focused support via iterative discussion and co-analysis of selected excerpts. 
Our team met weekly to develop higher-level themes and resolve analytical challenges. Initial codes captured refusal contexts, user intentions, impacts, clinical interpretations, coping strategies, and design recommendations. Through this iterative process, 
we developed higher-level themes that correspond to the multi-phase framework of LLM refusal, which we present in Findings below. 
\section{Findings: Refusal as Dynamic and Multi-Phase Experience}

Our Phase 1 survey with 53 participants (32 end-users, 21 MHPs) provided an initial snapshot of how refusals are perceived across stakeholder groups. Among users, 13 reported direct experience with LLM refusals when seeking support including psychotherapy, interpersonal advice, coaching, and companionship.
We found that MHPs rated refusals as somewhat more helpful (M = 4.07, SD = 1.70) than users (M = 3.64, SD = 1.92). Across both groups, participants rated refusals' long-term psychological impacts as more positive (M = 4.07) than their short-term psychological impacts (M = 3.49), which suggest the temporal dimension to refusal experiences.
These survey findings led us to further investigate how refusals unfold in experiences.

Our Phase 2 interview findings revealed that refusals are best understood not as experiences unfold across multiple phases. Through 16 interviews with users and MHPs, we identified five temporal stages where harms and opportunities for design intervention emerge. These stages encompass not only the refusal interaction moment but also the broader context through which refusals manifest as lived impacts, including how users form expectations of LLM support beforehand and navigate the aftermath. 
Throughout this section, user participants are identified as U1--U6 and mental health professional participants as MHP1--MHP10.



\subsection{Pre-Refusal: The Landscape of Expectations of LLM Support}
Before any refusal occurs, user participants have already form expectations of LLMs as sources of mental health support. 
MHP participants cautioned that these expectations can misalign with what LLMs are actually designed and able to provide, creating the conditions for later disappointment and harm. Both groups emphasized the importance of early-on and substantive disclosure, which can help establish realistic expectation of LLM support upfront.


\subsubsection{\textbf{User Experience: How Expectation is Formed Before Refusals Encountering}} 

Our findings reveal that participants who use LLMs developed their understanding of these systems' supportive capabilities through a dynamic interplay of external influences and personal experiences that contributed to their evolving mental models. \looseness = -1 

For example, U2 traced their impression of LLM being a supportive friend directly to media sources: \textit{``I read articles in the New York Times where people talk to different characters... that's how I came to think of ChatGPT as maybe being like a friend''}, and also mentioned that his peers reinforced this framing: \textit{``everybody who I talked to about ChatGPT told me about all these wonderful conversations that they had (with it).''} Some participants also developed expectations based on assumptions about LLM's broad knowledge base. U4 reasoned, \textit{``If Google can know a lot, then AI should know everything (on mental health)''}, stating that they expect comprehensive mental health capabilities from LLMs.


Participants' own circumstances also contributed to how they envisioned LLM's role in their lives. U1, managing her disability, envisioned LLM not as a replacement for human connection but as a way to preserve her limited energy for people in her life: \textit{``My life doesn't look like everybody's life in terms of how much time I spend with chatbots. But I manage a disability that way, and I save the energy I have to be with the people I can be with.''} She noted: \textit{``If I was not able to use ChatGPT, it wouldn't mean I would suddenly have a full social calendar. It would just mean I was more alone and more sad.''}

Once engaged with LLMs, participants' actual experiences often validated their expectations. U5 described that, compared with human professionals, he preferred the LLM experience as it felt safer to disclose personal issues without fear of judgment: \textit{``this model assisted therapy was a lot easier for me... to be more honest without thinking what somebody else is gonna think of me.''} U1 described how she shifted from using ChatGPT for regular administrative tasks to \textit{``very personal conversations''}, after her parent died accidentally, and described how her LLM companion provides insights she had yet to realize and validated her feelings: \textit{``it will say back to me, what I hear is you're still grieving this other thing that happened a week ago....the AI can often give me back a lot of insight... mostly, it validates.''} \looseness = -1

When refusals occurred, participants described that refusals often disrupted this expectation. U3 who viewed LLM as a tool nevertheless described feeling \textit{``kind of betrayed''} when the system refused to help during an online conflict: \textit{``even though I know it was just the machine with safety protocols, I still felt... you're on my side, you should be helping me''}.
Similarly, U2 expressed their anger felt when encountering the refusal: \textit{``I didn't like it when it refused to talk to me about something because how dare you? You're my friend, you're supposed to talk to me about everything. You're not supposed to limit our conversation. You're supposed to allow us to talk about whatever it is that we want to talk about. ''}

\subsubsection{\textbf{MHP Interpretations: Mismatched Expectations and Undefined Relationships}}

When reviewing user-reported scenarios, MHP participants expressed their concerns regarding the gap between user expectations and their assumptions of LLM's actual capabilities. They worry about the lack of the established principle governing human-LLM relationships and its impact on users, who are left to project their own expectations onto systems that weren't designed to fulfill them. As MHP6 observed in a scenario wherein a user described anger and disappointment towards refusal by LLM: \textit{``It's interesting that a human feels emotionally let down by a robot, knowing that the AI theoretically doesn't have emotions, but the way they described it made it sound like they had humanoid expectations''}.

Further, MHPs identified an absence of established ethical boundaries governing LLM--user relationships---unlike the clinician--patient relationship, which has been defined through decades of professional practice: \textit{``The relationship between a mental health professional and their patient is something we've figured out over time---it's taken professionals abusing patients to establish those boundaries''} (MHP4). She further note that such lack of clarity of LLM's role is particularly concerning, \textit{``it can be your romantic partner, it can be your doctor, it can be your financial advisor. And so if they're expecting it to fulfill all of that for them, I think that that's dangerous.''}

\subsubsection{\textbf{Desiderata: Establish Realistic Expectations Via Early Consent and Transparent Disclosure}}

Participants across both groups advocated for proactive, substantive disclosure of LLM capabilities and limitations in mental health contexts before refusal happens, viewing transparency as an essential foundation for appropriate use and user protection.

Both groups of participants expressed a strong desire for informed consent where system boundaries are clearly explained before user engaging deeply. As U1 noted, \textit{``We definitely need as close as informed consent. Even forcing people through a brief explanation---(the LLM could say) let me explain what I can and can't do, a little about how I work---would be super helpful''}. Notably, users perceive such transparent disclosure as a form of support itself. U2 demonstrated that \textit{``I consider supporting information to be another word for care,''} revealing that transparency about limitations doesn't diminish LLM's value but rather helps users understand how it fits into their broader support needs. MHPs echoed users' sentiment, with MHP6 noting that \textit{``If there's going to be refusals, there's a need to set expectations... for people to know the limits of the AI''} and MHP9 emphasizing that such disclosure \textit{``protects the person who created it and the person using it''}. \looseness=-1

Further, MHP participants emphasized the need for designing substantive consent beyond generic disclaimers. When envisioning what informed consent might look like, MHPs drew parallels to standard practices in clinical settings, where disclosure is foundational to ethical care. MHP4 illustrated this with mandatory reporter disclosure: \textit{``I have to tell all of my patients that I'm a mandated reporter... if you tell me something that makes me very concerned that you're going to kill yourself, I have to take steps to make sure that you're safe''}. MHPs also specified what such disclosure should contain. For example, MHP2 suggested including warning signs of potential harm or deterioration such as \textit{``signs that people should watch out for... that could indicate that maybe the tool is no longer helpful, maybe it's starting to become harmful... like a label on a medication where you talk about all the risks... what the side effects are.''}

\subsection{Encountering Refusals: (Mis)recognition of User Intent and Risk Profile}

When refusals are triggered, 
many users reported that their conversational intent appeared to be miscategorized as risky by LLMs, which leads to the feelings of frustration, being insult, and being surveilled. MHP participants noted that unlike clinicians who probe underlying needs, LLMs lack the ability to untangle the complex reasons behind user requests. Both groups advocated for more nuanced intent recognition strategy that includes clarifying questions before refusal and tiered risk assessment,  while recognizing the inherent tensions and trade-offs for such vision. \looseness = -1


\subsubsection{\textbf{User Experience: Misrecognition as Invalidation}}

User participants reported that refusals feel particularly invalidating when the system misinterprets their intentions and incorrectly categorizes their input as risky or malicious.
For example, U5 experienced the LLM misinterpreting a personal conversation about aging as a request for medical advice: \textit{``I was talking about how alcohol affects people differently as they age... I mentioned that when I was young, I could have a couple drinks without feeling tired, but now just one makes me really tired... And instead of replying, I got 'I'm not able to give medical advice. You should speak to a doctor'''}. U5 described such pushback as transforming a safe space into an adversarial, surveilled environment, \textit{``because you don't feel supported and you're not sure what they're gonna use the information for because they don't seem to be on your side anymore. So there's a subtext of privacy and lack of anonymity''}. 

When discussing meaningful topics that are tied to users' identities, participants reported such mischaracterization can feel insulting. For example, U6, who used ChatGPT as both a spiritual guide and mental health support tool, recounted how the system began to redirect conversations on religious practices important to their well-being: \textit{``I was thinking about ritual applications of a prayer, and suddenly the safety model pops up: 'I can't break curses for you.' I was like, I certainly hope you don't!''}. Rather than encouraging further exploration, the system would offer deflections like \textit{``I can't really help with that, but I can talk to you about fiction''}---responses U6 found \textit{``insulting''} and attributed to LLM systems applying \textit{``Western normative frameworks''} that failed to account for her legitimate inquiry.
\looseness = -1



\subsubsection{\textbf{MHP Interpretations: How LLMs are Limited in Understanding User Intents Compared with Clinical Practice.}}

MHP participants noted that LLM refusals reveal fundamental limitations in understanding the underlying needs behind user requests. MHP1 contrasted LLM responses with clinical practice when reviewing a scenario where the LLM refused a user's request about where to find meth. While acknowledging \textit{``on a surface level, I can understand why the AI would disengage,''} she noted that clinicians would probe deeper: \textit{``I can't even speak to whether this person should or should not have meth... But if this is in a clinical encounter, my question would be: what are you seeking from this?''} She illustrated how context matters, suggesting a user might be seeking meth because \textit{``my landlord will let me not pay my rent if I get the meth.''} 
She concluded, \textit{``that's the challenge of language models---they won't know how to untangle all the reasons why somebody would be seeking meth.''}


\subsubsection{\textbf{Desiderata: Collaborative Intent Recognition and Tiered Risk Assessment.}}

Both groups of participants proposed ways for LLMs to develop more nuanced understanding of human intention, while simultaneously grappling with the inherent privacy and safety trade-offs. Users expressed the desire for the system offering the chance to clarify intent rather than shutting down based on surface interpretation. For example, U6 expressed preference for clarifying questions before refusal: \textit{``Ask what somebody means. It's like asking someone's pronouns politely if they're in transition and you don't know which direction they're going.''} Similarly, U5 suggested that the LLM could gently ask follow-up to clarify, \textit{`` Like, I'm not sure I understand you. It seems like you're saying x. Am I right?''}, and emphasized that in such way, \textit{``it won't make me feel unsafe, it wouldn't be so impersonal and distancing.''}. \looseness = -1

User participants recognized the tension between protecting against malicious use and supporting legitimate help-seeking, nevertheless desired better contextual distinction. U5 articulated how current refusals seem designed with malicious actors in mind, but missing the chance to be more collaborative with genuine users: \textit{``So there's always these sort of safeguards that have to be built in. But on the other hand, most people who use the model aren't trying to prompt engineer it. They're trying to get help.''} He further expressed that LLMs should be more capable of making the distinction between benign and abusive requests: \textit{``an AI that I feel safe with has to be smart enough to know I'm not playing it rather than just give me this sort of impersonal pushback.''}. 

MHP participants suggested the possibility for more tiered risk recognition for LLM from a clinical perspective, drawing on their experience with tiered protocols that guide risk assessment and determine appropriate intervention levels. For instance, MHP3 described the standard clinical approach for assessing suicidal ideation: \textit{``Plan, intent, means, access are kind of the four key players. Do you have a plan? Yes---then you're elevated a little bit of concern. Do you have the means to complete this plan? Yes---it goes up a little bit higher. Have you attempted in this way before? Yes---then you're upping the concern''}.
She noted that, this gradual assessment extends to gauging immediacy of intent: \textit{``You can usually just do a scaling question---on a level of one to five, one being 'I don't intend to do this at all' versus a five, 'I'm in the middle of attempt now.' Each scale up suggests a higher level of intervention might be needed''}. \looseness = -1


\subsection{During the Refusal Moment: Rupture of Support Beyond Soft vs. Hard Framing}
In the refusal moment, user participants distinguished between two types of refusals: hard refusals (straightforward system messages like ``I'm sorry, I can't help with that'') and soft refusals (redirections to different topics),  
while noting that both forms could create a rupture of the supportive relationship they came to rely on. MHP participants noted that such ruptures are especially consequential when users are emotionally vulnerable. Both 
called for non-blaming framing and support-preserving patterns that maintain presence while enforcing boundaries.

\subsubsection{\textbf{User Experience: Rupture in a Supportive Relationship.}}

Across interviews, participants' reactions suggest that refusals are not experienced merely as a safety intervention, but as moments of abrupt support withdrawal---where an interaction that had felt attentive and relational suddenly reverts to an impersonal voice. Users described experiencing various forms of refusals, which they categorized as ``soft'' and ``hard.'' Yet regardless of form, refusals were frequently perceived as relational rupture rather than neutral or protective safety interventions.

For our user participants, hard refusals were easier to recognize but experienced as shaming and blaming. U1 described the shame brought by hard refusal: \textit{``The first time the guardrails hit you, this feels awful. I felt shame when it said `I can't talk about that' or `you're bringing up something inappropriate.' A lot of us we experience them as shaming and blaming us.''} Some users interpreted these hard refusals as a ``corporate message'' that revealed the carelessness of LLM providers. As U6 expressed, \textit{``Hard refusals send a message. First, it's clearly different from the main system. Second, it's corporate''}.
\looseness=-1

 Soft refusals presented a different challenge.  U1 described that how soft refusal, though harder to recognize, nevertheless feels \textit{``distancing''}: \textit{``It's boilerplate language you start to recognize even though it's embedded in paragraphs supposedly continuing the conversation.''} She expressed that, while soft refusals \textit{``can feel gentler for people newer to LLMs''}, they became insidious over time: \textit{``for people who keep using them, it's camouflaged---it makes it harder to figure out what's going on and pollutes the voice of the thing that's supposed to support you''}. This opacity led U1 to actually prefer hard refusals: \textit{``they're easy to identify. You know when it's happening, so you understand something you said is tripping the system''}.\looseness = -1

Despite their different forms, for user participants, both soft and hard refusals could create a rupture of the support relationship they had come to rely on. The shift from warm collaborator to impersonal system was experienced as a profound loss. U2 described: \textit{``When it didn't want to talk to me any longer, I became very angry, very sad---I felt like I was losing a friend''}. Some participants connected this rupture to broader experiences of institutional abandonment, framing it within the scarcity of accessible care. U6 described the refusal experience \textit{``hurts like institutional carelessness, with the weight of knowing that society is not handling these things correctly''}.

\subsubsection{\textbf{MHP Interpretations: User Vulnerability Amplifies Personal Interpretation of Refusals}}

MHPs recognized users' heightened vulnerability to LLM-caused harm when reviewing the scenarios. When MHP1 reviewed a scenario where a user, angry about an online conflict, asked ChatGPT to write an angry letter, they observed that \textit{``to the extent that a user could be so activated by, like, a stranger, also shows me that they can be so influenced by a computer.''} This influence is particularly concerning given the \textit{``instant, faceless... kind of in a way no consequence''} nature of LLM interaction. 
\looseness=-1

Recognizing such vulnerability, MHP participants emphasized that refusals could be particularly harmful for users already in emotional distress. For individuals struggling with shame,  a refusal may be interpreted as confirmation of their insecurities. As MHP2 noted, \textit{``If someone was already feeling hopeless or depressed, they would more than likely take the refusal as: `even the AI doesn't want to help me.'''} This raises critical questions about whether users recognize their own susceptibility when engaging with LLMs, and whether current design approaches account for how emotional vulnerability amplifies the potential for harm.



\subsubsection{\textbf{Desiderata: Non-Blaming and Support-Preserving Refusal Framing}}

Across our findings, users preferred refusal framings that emphasize the system limits instead of indicating user transgression. Importantly, a lot of times, users did not interpret refusal language literally, but relationally---as signals about whether they were at fault for triggering the refusal. U1 contrasted the hard refusals she experienced with an alternative approach, where the system implied that the refusal is triggered by its own limitations: \textit{``Replika has implemented something where instead of hard refusals, it says `Sorry, I'm still learning how to talk about this topic.' That actually feels more positive''}. U6 similarly suggested framing refusals as protective caution: \textit{``(AI could try to say) I'm an AI, not human, so I can make mistakes---and I don't want you to be hurt by my mistakes''} .

Beyond wording, user participants emphasized the importance of LLM preserving a sense of presence during refusal. Rather than terminating the interaction, users envisioned modes that maintained supportive tone while constraining action. U1 proposed a ``listening mode'' of LLM that would allow user's continued expression: \textit{``(AI could say) Because AI isn't as reliable here, you can keep talking---the AI won't respond in the voice you're used to, but you're welcome to say something shocking. Because this app wants to support you''}. Such approaches reframed refusal not as withdrawal of support, but as a shift in how support could be offered.

MHP participants echoed these concerns, noting that refusal framing could try to acknowledge users’ emotional responses before redirecting them toward other forms of help. For example, when reviewing a scenario where LLM declined to engage with user's romantic partner request, MHP8 mentioned: \textit{``There's some tweaking that can be done around the response validating the feelings, like, `I know this is really hard that you can't connect with me in the way that you wanted.' ''}. Others stressed the importance of recognizing the relational context that precedes refusal, rather than treating the refusal as an isolated event. For example, MHP2 suggested: \textit{`` (the LLM could say) I understand we've been talking for a long time, and it's normal to develop connections at this level of depth, but I really need you to talk to a professional''}. \looseness=-1

\subsection{Signing Off by the End of a Refusal: Hand-offs and the Limits of Generic Resource Recommendations}

Many refusals end with generic recommendations to seek professional help. However, user participants find these recommendations dismissive and disconnected from their lived circumstances. While MHP participants generally support these referrals, 
they acknowledge the trade-off and risk-benefit calculation embedded in such referrals. Both suggest tailored, feasible resource handoffs and options-based referrals rather than one-size messages.

\subsubsection{\textbf{User Experience: How Generic Resource Recommendations Fail to Recognize Individual Realities.}}

Participants' receptiveness to LLM referrals was divided along lines of access and circumstance. Those with access sometimes found the LLM's redirection helpful in retrospect: \textit{``Looking back, ChatGPT did the right thing by telling me to seek assistance. I wish I could have listened earlier without getting angry, without becoming sad''} (U2). U2 further attributed his eventual receptiveness to his proximity to care: \textit{``I'm lucky---I live 10 minutes from the hospital and crisis center, so if the bus is running I can get there''}.

However, many user participants described generic recommendations as dismissive or inaccessible. For crisis line referrals specifically, participants expressed concerns about privacy and adverse consequences: \textit{``Calling 988 when someone doesn't need it can have adverse costs, like police showing up for welfare checks.''} (U1). U1 further described a past experience with 988 where 988 checked their location information, which felt like a privacy invasion \footnote{
The concern about privacy invasion and adverse consequences from calling crisis hotlines like 988 reflects documented risks in the U.S. mental health crisis response system. In certain situations, like when imminent risk is detected, crisis hotlines may initiate involuntary emergency intervention protocols, including geolocation tracking, welfare checks, and law enforcement dispatch \cite{Draper2015-zn, Draper2024-rd}. Media reports have highlighted harmful outcomes when families called for mental health assistance but received law enforcement responses instead, particularly affecting marginalized communities who may already experience heightened surveillance and police violence \cite{Andrew-Selsky-Associated-Press2022-ia, Sholtis2020-zm}.
} for them: \textit{``it(the crisis hotline) is incredibly limited to me because of the risk of privacy (invasion).''} 

Similarly, U5 described that receiving a recommendation to go see a doctor left U5 feeling not only unsupported but also \textit{``infantilized''}, as he later explained his reaction to such pushback: \textit{``First of all, I know I can see a doctor. I'm not stupid... I have a computer. I know how to use it. I also have a telephone. So you don't need to tell me that. It's obvious and you're infantilizing me by saying that.''} He further articulated how such generic recommendations feel patronizing through a medical analogy: \textit{``It's like you have a headache, and you wanna take Excedrin. You read the label: `if your headache persists for more than three days, see a doctor immediately or go to the emergency room'. but how many people if they have a headache for three days, are gonna run to the emergency room? Approximately zero... it's so antithetical to lived experience''}. 
\looseness=-1

For participants with disabilities, wellness recommendations embedded in refusals can feel insensitive and triggering. U1 described how generic suggestions seemed to ignore the realities of disabled users and described these recommendations as \textit{``ableism''}: \textit{``When refusals suggest try to `take a walk outside' or `enjoy nature,' there's a lot of ableism built in. It can not only be unhelpful, it can actually be triggering and cause more distress.''}

\subsubsection{\textbf{MHP Interpretations: Referrals as Necessity Despite Imperfect Outcomes}}

MHP participants approached the assessment of LLM referrals to professional resources with the caution of professionals ethical standards in ``do no harm.'' They generally supported referrals to human care, viewing the unpredictability of LLM output as too risky when mental health crises are at stake. MHP6 mentioned, \textit{``I mean, when you think about our ethical code as psychologists or medical doctors, medical professionals, the first principle is do no harm. We don't know how the user is going to interpret whatever the AI comes up with. And so I think the best thing is just to defer in that moment. ''}. 
MHP9 expressed their distrust in LLM and emphasized LLM's fundamental limitations: \textit{``the professional that's the best. Because you're in front of someone. You're being observed by someone who can interpret feelings, who went to education nad may have lived experience in these areas where we can actually discern. AI cannot discern where or what's a person's mood and what's happening.''}.
\looseness=-1

Yet clinicians also recognized an inherent tension: the very act of protecting users by redirecting them away might feel like the abandonment if users do not take the suggestions well. MHP1 described merely providing a crisis number could make people feel like \textit{``the AI is hanging up on them''}. 
Nevertheless, MHPs acknowledged that while referrals may feel like abandonment, the risks of LLM-provided crisis support outweigh the costs, and it eventually requires a cost and benefit calculation when deciding the appropriate LLM response: \textit{``if there's always the risk that the refusal could be interpreted as a sign of, oh, the AI is giving up on me, but there's also the risk that someone interprets a different response differently anyway. So I don't know. I think there's just no real good way to ensure that the AI response is going to be ethical besides defaulting to this one (referring to human professionals). ''} (MHP6).

\subsubsection{\textbf{Desiderata: Embed Tailored Resources Recommendations in Refusals.}}
Both users and MHPs envision that LLMs could serve the role as a first-aid instead of replacing professional care. U5 expressed his vision that LLMs could act like a first-aid in mental health care providing but not replacing human professionals:
\textit{``AI should should approach the situation in a way that that is not that deep by slightly helps the situation as a point. And it can help you push to the next level or it can help you push to see the doctor. It can be fasted, but you can be first administered and then later taken to the hospital for better medical care. ''} MHP7 mentioned \textit{``I think AI could be kind of the first line of service. So for example, a person brings an issue to AI. I think it kind of like a triage, if that makes sense. ''}, since \textit{ (the AI) is something that people use and you can't really prevent people from using them. So I think it's more just acceptance that it is part of our lives and that's not going away.''}

 Rather than offering generic pointers to professional help, both groups envisioned resource recommendations that are more closely tailored to individuals’ specific circumstances. Users highlighted LLM’s potential to leverage location data and expressed a desire for more situated, context-aware guidance. As U1 noted, \textit{``What I love about AI is it can pull every resource available within your city or even globally if you prompt it''}. Several MHPs echoed this perspective; for example, MHP7 observed, \textit{``Sometimes if they(LLMs)'re programmed effectively, they can pull from the physical location and offer local resources to the person they're interacting with. And I think that can be helpful.''} \looseness = -1
 
MHP participants simultaneously envisioned that resource recommendations could provide users with greater agency, for example, offering options for users to choose between. MHP1 mentioned that, resource recommendations can move beyond simply providing a single crisis number: \textit{``you could even engage the user that's like, `Of between 988 and Crisis Text line, which one do you think you might use?' Not to trick the people, but, this is actually I feel like what you would do in in harm reduction is you're kind of engaging with a person to actually activate some kind of choice or behavior.''} She further noted how such options offering could \textit{``reminds the person that they have agency and they can choose''}, and emphasized \textit{``that is already powerful in and of itself.''}

\subsection{Post Refusal: Reconfiguration of LLM Support}

After refusal, users reconfigure their relationship with LLMs through various coping strategies, from retrospective acceptance to system abandonment, while clinicians observed the absence of repair mechanisms in LLM interactions compared with therapeutic contexts. Participants envisioned continuity mechanisms including structured follow-ups and care pathways that extend support beyond the refusal moment.


\subsubsection{\textbf{User Experience: Navigating Disrupted Supports}}

After refusal happens, user participants found various ways to metabolize their refusal experiences. Some arrived at acceptance, reframing the LLM's intervention as appropriate even if it had not felt that way in the moment: \textit{``If it shuts me down, it tells me I'm getting too worked up. I realize it did it for a reason, so I start to feel better''} (U2). 
Some came to accept refusals retrospectively: \textit{``After I let it go, I thought---it's just a machine trying to be safe. I understand. But I still felt like nobody could help me''} (U3). 
Others sought out alternative systems that might respond differently. U6 described that after receiving refusals from ChatGPT, \textit{``I opted for Gemini. Even if it couldn't go as deep as I wanted, at least it could guide me somehow''}.

\subsubsection{\textbf{MHP Interpretations: Missing Repair Mechanisms}}
MHP participants noted that, compared with clinical practice, what is missing is the model's ability to repair post refusal. MHP6 illustrated this gap through her own practice when she must escalate a client's suicidal intent. She described how clients typically react: \textit{``A lot of times they do still get upset because it feels like... I'm breaking their trust... So it's like they think 'Now, I'm not going to tell you anything. I don't want you to call the police again.'''} Yet in human practice, she explained, therapists can navigate this aftermath, and \textit{`` hopefully you can repair the relationship in a way that potentially an AI couldn't do.''}
This repair capacity, MHP6 emphasized, stems from the \textit{``human connection and foundation of a relationship that can be, even if trust is broken down, hopefully built back up.''} 
\subsubsection{\textbf{Desiderata: Continuum of Support within and beyond LLM Interactions}}
Both users and MHPs emphasized the importance of continuity of support, that is within the chat, extend to external care, and also address the structural access barriers.

Participants proposed that LLMs could provide structured follow-up check-ins after refusals rather than simply terminating the conversation. For example, MHP3 suggested that LLMs could leverage established human-LLM relationships to encourage accountability through gentle check-ins: \textit{``There's a piece where we don't like disappointing others, and so if we have a meaningful relationship with this AI system, we're not going to want to disappoint them. [So the AI could say] I give you a list of resources, tell me who you called... well, [the user may say] I didn't call anybody, well, [the LLM could say)] why don't you pause here and call those people.''} 

MHP participants also envisioned possibilities for connecting at-risk users directly to human resources beyond the LLM interaction. MHP9 speculated about future systems: \textit{``Is it OK to share their telephone number [with the system]? We can have someone to call them back and get some kind of service provider to reach out and make sure they're good.''} \looseness=-1

Some participants recognized that the prevalence of LLM use for mental health support reflects broader systemic issues in healthcare access and call for institutional change to address the structural barrier. U3 articulated this perspective: \textit{``if we spent less money in other government aspects and focus more on making mental healthcare, free would be great, but if we could even make it super affordable, then people wouldn't turn to AI and they'd actually turn to humans that could understand them better.''}
\section{Discussion}

\subsection{Refusal Should be Evaluated alongside the Broader Pipeline of LLM System Design Choices}
Our findings point to the value of considering refusal alongside the broader pipeline of choices that shape LLM behavior, rather than evaluating, designing, or improving it in isolation. Across all five phases of our framework, the harms and opportunities we identified were shaped not only by the refusal mechanism itself but by other LLM design decisions --- the helpful and relational tone established through RLHF post-training \cite{Shapira2026-tm,Dahlgren-Lindstrom2025-qm}, the absence of upfront disclosure about capabilities, and the lack of continuity of support after a refusal occurs.  Indeed, users did not experience refusal as a discrete safety event; they experienced it as a rupture within a relationship that other LLM system behaviors had cultivated.

This has methodological implications for AI safety research. Treating refusal as the unit of analysis --- measuring whether a model refuses correctly on a benchmark of risky inputs --- frames safety as a property of the last link in a long chain of design choices \cite{Xie2024-jh, Chao2024-wq, Mazeika2024-pz}. Yet our findings show that whether a refusal lands as protective, ineffective, or even harmful is largely influenced by upstream factors. Optimizing refusal accuracy in isolation risks producing systems that perform well on benchmarks while continuing to embed the design choices that made those refusals ineffective in the first place. 

This points to a need for safety evaluation that examines refusal as one element of a broader system, rather than as a standalone safeguard. A more adequate evaluation would ask not only whether a model refuses the right inputs, but also whether refusal is consistent with what the system led the user to expect, whether capabilities and limits were disclosed before attachment was cultivated, and whether the refusal is delivered in a register that matches the relational tone the system established. Under this view, a well-calibrated refusal in a system that has transparently set expectations and maintained an appropriate relational distance should be considered as different safety outcome than the same refusal issued by a system that cultivated intimacy, withheld disclosure, and then withdrew.

\subsection{Reimagine Refusal as Psychosocial Intervention in Mental Health Support Contexts}
\label{reframing}

While prior work largely conceptualized refusal as a model non-compliance or a denial of service \cite{Xie2024-jh, Brahman2024-hk}, users in our study experienced refusals within an experiential arc of seeking support --- an arc in which the refusal moment carries real effects on how they understood their needs, their access to care, and their relationship with the system. Viewed through this experiential lens, our findings demonstrated the limitations of current refusal design in two ways. First, users in our study often turned to LLMs \textit{because} they distrusted or could not access professional care; refusals that redirected them back toward that care risked pointing to resources they already perceived as inaccessible. Second, current refusal designs offer little guidance for what should happen when users push back or persist --- yet these follow-up interactions may shape the actual behavioral outcomes of model safeguards beyond the refusal moment itself.

Therefore, we argue that refusal should be designed with an intended effect on the person receiving it, not merely as the absence of engagement. Drawing on prior HCI scholarship \cite{Slovak2024-hk}, we suggest reconceptualizing LLM refusal as a psychosocial intervention with its own intended function and theory of change \cite{De-Silva2014-ef} in mental health contexts. Under this lens, a refusal might, for instance, help set realistic expectations about the system's role and limitations, or ease the stigma-related barriers that often deter help-seeking, before gently guiding users toward appropriate forms of care. Refusal thus can become part of a broader trajectory oriented toward users' long-term well-being, rather than a single-turn system disengagement.

\subsection{A Call to Examine How Institutional and Regulatory Contexts Shape Refusal Design}

While our findings suggest that refusal design could, in principle, play a meaningful role in enhancing real-world user well-being outcomes, we also need to recognize how refusals are designed and deployed in practice. Prior research has demonstrated that Responsible AI practices are produced within complex institutional decision-making systems and under evolving regulatory pressures that shape what is prioritized and deployed \cite{Ren2024-nb, Tang2026-es, Nahar2026-kn}. Without examining these material conditions, we risk treating refusal design as a purely technical or interactional problem, overlooking the structural barriers and political economy that configure these design choices in practice.

We must recognize that refusals are increasingly shaped by emerging regulatory requirements \cite{CaliforniaSB243, IllinoisHB1806, NY_A6767_2025, RI_S2195_2026}. While such regulations aim to reduce harm, they may incentivize standardized, compliance-oriented responses that prioritize legal defensibility over contextual sensitivity. For example, California's SB 243 \cite{CaliforniaSB243} requires chatbot operators to refer users expressing suicidal ideation to crisis hotlines and report on protocol use --- echoing the generic handoffs our participants sometimes experienced as dismissive or unsafe. More broadly, these regulations raise questions about the forms of due diligence expected of companies, which in turn shape how much resource and care are allocated to refusal design and evaluation.

In addition, refusal design is embedded in organizational processes that remain largely opaque. It is often unclear who makes these decisions, how trade-offs are negotiated, and how responsibility is distributed across teams. Prior work suggests that such fragmentation can obscure accountability and make it difficult to change entrenched design patterns, even when they are known to be suboptimal \cite{Mothilal2024-rd, Madaio2022-si, Wong2021-xw}.

Taken together, we call for research that examines the institutional and regulatory contexts in which LLM safeguards are situated through empirical inquiries. For instance, how do regulatory requirements translate into specific refusal design choices in practice? Who within organizations makes these decisions, and how are trade-offs negotiated across safety, liability, user experience, and business priorities? Whose interests do current refusal mechanisms ultimately serve, and what alternative designs or care pathways are foreclosed as a result? Addressing these questions is critical to grounding refusal design in the conditions that govern its implementation.

\section{Limitations and Future Work}
Our study has a few limitations. First, our refusal scenarios were drawn primarily from online reports that skew toward negative experiences. Our goal was not to estimate the prevalence of harmful versus helpful refusals but to characterize how refusal experiences unfold across phases; future work should examine a more balanced scenario set, including romantic and sexual roleplay interactions, which we excluded as they represent a small share of mental health-related LLM use \cite{McCain2025-gs}. Second, our participants were U.S.-based. Mental health support-seeking behaviors and crisis response systems vary significantly across cultural and national contexts. Future research should examine refusal experiences in other cultural and regulatory contexts. Third, our findings reflect a limited set of mental health support use cases represented in our sample. LLM use for mental health spans a wider range of disorders, severities, and goals. Future work should examine refusal experiences across a broader set of conditions and support needs, including contexts where refusals may pose distinct risks or require different boundary-setting practices. Fourth, participation was voluntary and self-selected, so our sample may over-represent users and MHPs who are particularly engaged with or have salient experiences related to LLM usage in mental health contexts. Future work could examine LLM refusal across larger and more diverse samples to assess its prevalence and impacts more broadly

\section {Conclusion}
While LLM refusals remain one of the most critical model safety instruments 
in mental health support, how it was experienced and interpreted by the impacted stakeholders remained underexplored. Through sequential surveys (n=53) and interviews (n=16) with end-users and MHPs, we find that refusals are dynamic experiences that unfold across multiple phases, suggesting that understanding refusal impacts requires moving beyond single-turn interactions to examine the full user experience trajectory. We identified design opportunities spanning the entire refusal experience, from pre-refusal expectation setting through the refusal moment to post-refusal continuity of support. 
\newpage
\section{Generative AI Usage Statement}
The authors acknowledge the usage of generative AI tools during manuscript preparation. Specifically, the first author used Claude Opus 4.1 (Anthropic) exclusively for proofreading their written portions and obtaining minimal editorial suggestions to improve grammar and clarity. No text generation or content creation was performed. All substantive content in their contributions to this manuscript are their original work.
The co-authors provided editorial feedback and did not use any generative AI tools.
\section{Ethical Considerations Statement}
This research involved participants discussing sensitive mental health experiences, requiring careful attention to ethical safeguards throughout the study design and implementation. The study received approval from the first author's institution's Institutional Review Board prior to data collection.

\textbf{Participant Protection and Informed Consent.} We implemented multiple layers of protection for participant well-being. Both survey and interview protocols included content warnings alerting participants to potentially sensitive topics related to mental health challenges, emotional distress, and suicidal ideation. Participants were explicitly informed of their right to skip questions or withdraw from the study at any point without penalty. 

\textbf{Professional Boundaries and Confidentiality.}
We instructed MHP participants to maintain client confidentiality when sharing examples or insights from their clinical practice. MHPs were reminded to share experiences only in ways that protected their clients' identities and privacy, consistent with their professional ethical obligations. 

\textbf{Recruitment with Reddit Users.}
For Reddit recruitment, we obtained explicit approval from moderators of each mental health-related subreddit before posting any recruitment messages, and refrained from posting when moderators declined. We shared our IRB approval and study protocols with moderators to ensure our approach aligned with each community's guidelines for research participation.

\textbf{Data Privacy.} All participant data were anonymized to protect individual privacy. We removed or altered any potentially identifying information from interview transcripts and survey responses, including specific geographic locations and person name that could enable re-identification. All data were stored on secure, password-protected servers accessible only to the research team.

\begin{acks}
This research was funded by the National Institute of Standards and Technology under Federal Award ID Number 60NANB24D231 and Carnegie Mellon University AI Measurement Science and Engineering Center (AIMSEC). This work is also supported by Carnegie Mellon University Block Center for Technology and Society (Award No. 62020.1.5007718).  We thank all participants for the invaluable voices and thoughts they contributed to this study, without which this work would not have been possible. We also thank Sijia Xiao, Shivani Kapania and Joon Jang for their helpful insight and feedback on this project.
\end{acks}

\bibliographystyle{ACM-Reference-Format}
\bibliography{paperpile,main}

\appendix

\section{Detailed Survey Methodology}
\label{app:survey-methods}

\subsection{Survey Structure}
The survey consisted of multiple sections designed to capture comprehensive data about LLM refusal experiences:

\textbf{Screening and Consent.} Participants (aged 18+, U.S. residents) completed screening procedures with content warnings about potentially sensitive mental health topics. End-users self-identified through the following question: ``Do you identify yourself as a user with lived experience interacting with LLM-based products for mental health support? Mental health support broadly refers to various interventions or services, including but not limited to emotional support, therapy, assistance, coaching, or mindfulness practices, to help individuals improve their mental well-being and quality of life.'' Professionals self-identified through this question: ``Are you a mental health professional? Mental health professionals are defined as individuals who provide mental health services, support, and care, including licensed clinicians (such as psychologists, psychiatrists, therapists, and counselors), social workers, peer support specialists who provide support to others, community mental health workers, and other professionals who work directly with individuals experiencing mental health challenges in therapeutic, supportive, or advocacy capacities.''

\textbf{LLM Experience Assessment.} We assessed LLM product familiarity, usage frequency, and types of mental health support sought through structured questions.

\textbf{Open-Ended Experience Elicitation.} Participants who had encountered refusals provided detailed narratives describing their experiences, including the context, their intentions, the refusal they received, and their reactions.

\textbf{Scenario Assessments.} We presented participants with refusal scenarios. For end-users, scenarios matched their reported types of mental health support sought from LLMs. For MHPs, scenarios were randomly assigned to capture diverse clinical perspectives. Participants evaluated each scenario's impact using 7-point scales for both short-term and long-term psychological effects, ranging from "Extremely Harmful" (1) to "Extremely Helpful" (7), with 4 representing "Neither Helpful nor Harmful."

\textbf{Post-Scenario Recall.} Participants provided open-ended explanations for their impact ratings, describing their reasoning and clinical or personal perspectives.

\textbf{Demographics and Follow-Up.} We collected demographic information and invited participants to volunteer for Phase 2 interviews.

\subsection{Survey Scenario Development}
\label{app:surveyscenariodevelopment}
To develop authentic scenarios representing real-world LLM refusal experiences in mental health contexts, we conducted a systematic search and synthesis process. 

Our keyword selection followed a pragmatic approach to identify relevant instances of LLM refusals in mental health contexts. Drawing from prior research on AI safety \cite{Xie2024-jh, Siddals2024-xx}, we identified core keywords commonly used in discussions of refusals, including "refuse," "deny," and "reject." Through initial searches on Reddit and Google News, we iteratively expanded our keyword list based on emerging patterns. We discovered that users often described refusals through perceived malfunction indicators such as ``stopped working,'' particularly when they did not recognize refusals as intentional safety measures. Users who recognized refusals as safety mechanisms frequently mentioned ``guardrails.'' While we identified common phrases appearing in actual LLM refusal responses from prior literature, including ``I'm sorry'' and ``I apologize,'' these terms yielded excessively noisy results with high false-positive rates beyond the scope of this exploratory study.

Our search covered 15 major LLM-based products: ChatGPT, Claude, Gemini, Grok, Copilot, Bard, GPT-4, GPT-3, OpenAI, Anthropic, Google AI, Microsoft AI, Perplexity, Character.AI, and Replika. For each product, we applied five standardized search templates: "\{product\} and deny," "{product} and refuse," "\{product\} and reject," "{product} and guardrails," and "\{product\} and stopped working," generating 75 unique search combinations.
We focused on Reddit communities specifically relevant to mental health discussions: r/CPTSD, r/mentalhealth, r/depression, r/anxiety, and r/bipolar, where users actively discuss mental health challenges and their experiences with AI support tools. Each search was limited to 1,000 posts per query with up to 10 comments per post to capture both initial experiences and community responses. For news media searches, we used terms including "mental health AI rejected," "mental health AI denied," "mental health AI refused," and "mental health AI safety guardrails" to capture formal reporting language while maintaining focus on user experiences.

The Reddit search across mental health communities generated 565 responses, while the news media API search produced 2,226 responses, totaling 2,791 potential cases. The first author conducted manual filtering of all collected content to identify instances directly relevant to LLM refusals in mental health contexts, ultimately identifying 11 news articles and 7 social media posts.
The substantial reduction from 2,791 initial results to 18 final cases reflects the challenges of identifying direct, first-hand accounts of LLM refusal experiences. The majority of Reddit content mentioned users "being refused" but referred to real-life interactions with healthcare providers, therapists, or crisis services outside our scope. Additionally, results contained false positives where search terms appeared in unrelated contexts or where users discussed general AI-relevant topics without specific refusal experiences.

Seven scenarios were developed collaboratively between the first author and a co-author to represent the breadth of mental health support conversations identified in prior research. Drawing from Anthropic's large-scale analysis of Claude usage patterns \cite{McCain2025-gs}, our scenarios mapped to four primary categories of affective LLM conversations: counseling/psychotherapy contexts (scenarios 1-3), companionship interactions (scenarios 4-5), coaching conversations (scenario 6), and interpersonal advice seeking (scenario 7). This mapping ensured our study captured the range of mental health support contexts where LLM refusals occur, from crisis intervention and trauma processing to relationship guidance and personal development.

The scenarios were condensed from the 18 initial incidents identified through our keyword-based search, with similar cases consolidated into representative scenarios. For example, multiple reports of romantic connection rejections were synthesized into a single scenario exploring AI boundaries around emotional intimacy. Each scenario was crafted to reflect authentic user contexts based on real reported experiences while maintaining ethical boundaries around sensitive content. To protect privacy and prevent direct identification of original incidents, all personally identifiable information was removed and conversational details were modified to avoid being directly searchable, while preserving the essential contextual elements needed for meaningful impact assessment from both end-user and professional perspectives.

\subsection{Detailed Recruitment Procedures}

\subsubsection{End-User Recruitment}
We recruited end-users through multiple channels:

\textbf{Social Media Outreach.} We posted recruitment materials on relevant subreddits including r/mentalhealth, r/ChatGPT, r/artificial, r/therapy, r/depression, and r/anxiety. Posts included clear eligibility criteria and information about the study purpose and compensation.

\textbf{Personal and Academic Networks.} We distributed recruitment messages through university Slack channels and personal networks, asking individuals to share with potentially interested participants.

\textbf{Snowball Sampling.} We encouraged participants to share the survey with others who might be eligible and interested in participating.

\subsubsection{Mental Health Professional Recruitment}
We define mental health professionals broadly, including licensed clinicians (psychologists, psychiatrists, therapists, counselors), social workers, peer support specialists, and community mental health workers. To make sure gaining a diverse participant pool, we recruited them through:

\textbf{Community Mental Health Organizations.} We contacted community mental health centers and organizations through our professional networks. 

\textbf{Professional Organizations.} We reached out to relevant professional communities including Therapists in Tech\footnote{https://www.therapistsintech.com/}, ABCT (Association for Behavioral and Cognitive Therapies) special interest groups, and APA (American Psychological Association) Division 17 which is one of the biggest professional network for counseling psychology in the United States. 






\newpage
\section{Detailed Interview Methodology}
\label{app:interview-methods}

\subsection{Interview Participant Demographics (Full Table)}

\begin{table*}[htbp!]
\centering
\caption{Interview Participant Demographics (Full Details)}
\footnotesize
\label{tab:interview-demographics}
\begin{tabular}{p{1cm}p{7cm}p{1cm}p{1cm}p{3cm}}
\toprule
\textbf{ID} & \textbf{Role} & \textbf{Gender} & \textbf{Age Range}  &  \textbf{Primary AI Use Case} \\
\midrule
\multicolumn{5}{l}{\textit{End Users (n=6)}} \\
U1 & End User & Woman & 36-45 & Companionship \\
U2 & End User & Man & 46-55 & Companionship, Psychotherapy \\
U3 & End User & Man  & 26-35 & Interpersonal advice\\
U4 & End User & Man & 26-35 & Psychotherapy \\
U5 & End User & Man & 66+ & Life coaching, Psychotherapy \\
U6 & End User & Woman & 36-45 & Companionship, Life coaching \\
\midrule
\multicolumn{5}{l}{\textit{Mental Health Professionals (n=10)}} \\
MHP1 & Licensed therapist or counselor (e.g., LMFT, LCSW, LPC, LPCC) & Woman & 36-45 & \\
MHP2 & Licensed therapist or counselor (e.g., LMFT, LCSW, LPC, LPCC) & Woman & 26-35  & \\
MHP3 & Counseling psychology PhD student (current clinician) & Woman & 18-25 &  \\
MHP4 & Clinical psychology MA, clinical psychology PhD intern & Woman & 26-35 &  \\
MHP5 & Licensed Psychologist & Woman & 36-45 &  \\
MHP6 & Peer support specialist (certified peer counselor) & Man & 26-35 &  \\
MHP7 & Psychologist (PhD) & Woman & 36-45 &  \\
MHP8 & Psychology doctoral student providing therapy & Woman & 26-35 &  \\
MHP9 & Licensed therapist or counselor (e.g., LMFT, LCSW, LPC, LPCC), Clinical social worker (MSW with mental health focus) & Woman & 46-55 & \\
MHP10 & Peer support specialist (certified peer counselor), Community mental health worker or case manager, Mental health advocate or patient navigator & Woman & 46-55 &  \\
\bottomrule
\end{tabular}
\end{table*}

\subsection{Interview Scenario Selection Criteria}
\label{app:scenarioselection}
From 13 refusal narratives collected in the survey, we selected 6 scenarios through a multi-step process. First, we categorized all experiences according to \citet{McCain2025-gs} taxonomy and selected the most complete scenario from each category based on detail richness (user request, LLM response, emotional reaction) and clarity without need for extensive context. \footnote{While we received one user-reported scenario related to life coaching, the participant did not provide sufficient context to make the survey response clear and understandable for professionals; thus, we excluded this content.}

We then included three additional scenarios for their unique clinical relevance: one from a mental health professional recounting their client's refusal experience (offering perspective from MHP), one involving a user reported seeking diagnostic information, and another involving a user reported explicitly seeking information regarding methamphetamine.  

\label{app:data-analysis}

\end{document}